\documentclass[prl,reprint,superscriptaddress,tightenlines,twocolumn,showpacs,floatfix,longbibliography,footinbib]{revtex4-2}

\pdfoutput=1

\usepackage{latexsym,amsmath,amsfonts,amssymb,bm,empheq}   
\usepackage{nicefrac} 
\usepackage[dvipsnames]{xcolor}
\usepackage{tabularx}   
\usepackage{booktabs}
\usepackage{multirow}
\usepackage{array}
\usepackage{tabularx}
\usepackage{subfigure}  
\usepackage[dvipsnames]{xcolor}
\usepackage{graphicx}
 \usepackage{relsize}

\definecolor{nblue}{RGB}{28,130,185}
\definecolor{cgreen}{RGB}{76,153,0}
\definecolor{myorange}{RGB}{245,156,74}

\usepackage{hyperref}
\hypersetup{colorlinks=true, citecolor=magenta, urlcolor=-myorange}

\newcommand{\dif}{\mathrm{d}}
\newcommand{\calE}{\mathcal{E}} 
\newcommand{\calR}{\mathcal{R}} 
\newcommand{\calA}{\mathcal{A}} 
\newcommand{\ellb}{\ell_{\rm B} } 
\newcommand{\Li}{\mathrm{Li}} 

\newcommand{\calY}{\mathcal{Y}}

\newcommand{\sm}[1]{\textcolor{black}{#1}}

\graphicspath{{Plots/}}

\begin{document}

\title{Long-range fluctuation--induced forces in driven electrolytes}
%
%
\author{Saeed Mahdisoltani}
\affiliation{Rudolf Peierls Centre for Theoretical Physics, University of Oxford, Oxford OX1 3PU, United Kingdom}
\affiliation{Max Planck Institute for Dynamics and Self-Organization (MPIDS), D-37077 G\"ottingen, Germany}
%
%
\author{Ramin Golestanian}
\email{ramin.golestanian@ds.mpg.de}
\affiliation{Max Planck Institute for Dynamics and Self-Organization (MPIDS), D-37077 G\"ottingen, Germany}
\affiliation{Rudolf Peierls Centre for Theoretical Physics, University of Oxford, Oxford OX1 3PU, United Kingdom}

\date{\today}
\begin{abstract}
We study the stochastic dynamics of an electrolyte driven by a uniform external electric field and show that it exhibits generic scale invariance despite the presence of Debye screening. The resulting long-range correlations give rise to a Casimir-like fluctuation--induced force between neutral boundaries that confine the ions; this force is controlled by the external electric field, and it can be both attractive and repulsive with similar boundary conditions, \sm{unlike other long-range fluctuation--induced forces.} This work highlights the importance of nonequilibrium correlations in electrolytes 
and shows how they can be used to tune interactions between uncharged biological or synthetic structures at large separations. 
\end{abstract}

\maketitle

%
%
%
Fluctuation--induced forces (FIFs) can arise in a wide range of systems where external objects modify the spectrum of the  fluctuations in a correlated medium~\cite{kardar99friction,gambassi2009review}. 
Such forces only act at short distances when the confined  fluctuations have a finite correlation length, for instance the Debye screening length in electrolytes~\cite{jancovici2004screening,lee2018casimir}; scale-free correlations, however, can give rise to long-ranged FIFs with universal properties~\cite{RMP2010}, e.g., in the case of Casimir attraction between metallic plates in vacuum~\cite{casimir1948attraction} and forces arising from critical fluctuations in thermal equilibrium~\cite{fisher1978wall,gambassinature}. 
The extent to which critical Casimir forces can be controlled 
have especially been investigated in recent years due to their   practicality in colloidal systems~\cite{maciolekreview}.
%
Out of thermal equilibrium, long-range correlations are common as they arise, e.g., from the interplay between the  conservation laws 
and a mismatch between fluctuations and dissipation \cite{garrido90conservative,grinstein90conservation,hwa89dissipative}. The ensuing FIFs have been studied in a variety of 
settings such as nonuniform temperature profiles \cite{najafi2004soret}, nonequilibrium diffusive dynamics \cite{aminov2015neqfif}, temperature quenches \cite{rohwer2017transfif,rohwer2018neqfif}, active systems \cite{activecasimir}, and Brownian and driven charged particles \cite{deanbrownian2014, deannoneqtune2016}. 

\begin{figure}[b]
	\centering
\hskip-1cm
	\begin{minipage}[c]{.42\linewidth}
		\centering
		\includegraphics[width=\linewidth]{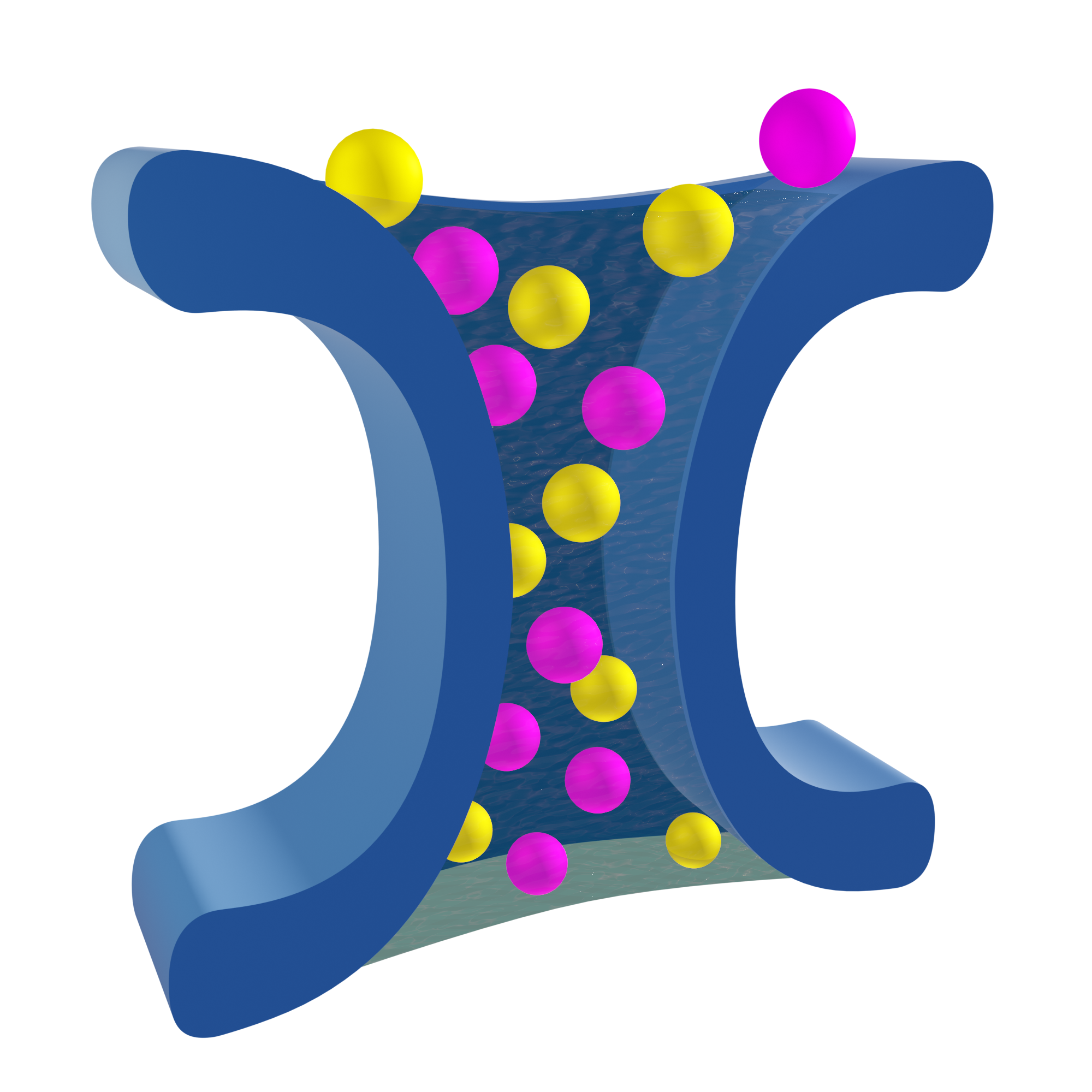} 
	\end{minipage} 
	\hskip.25cm
	\begin{minipage}[c]{0.42\linewidth}
		\centering
		\includegraphics[width=\linewidth]{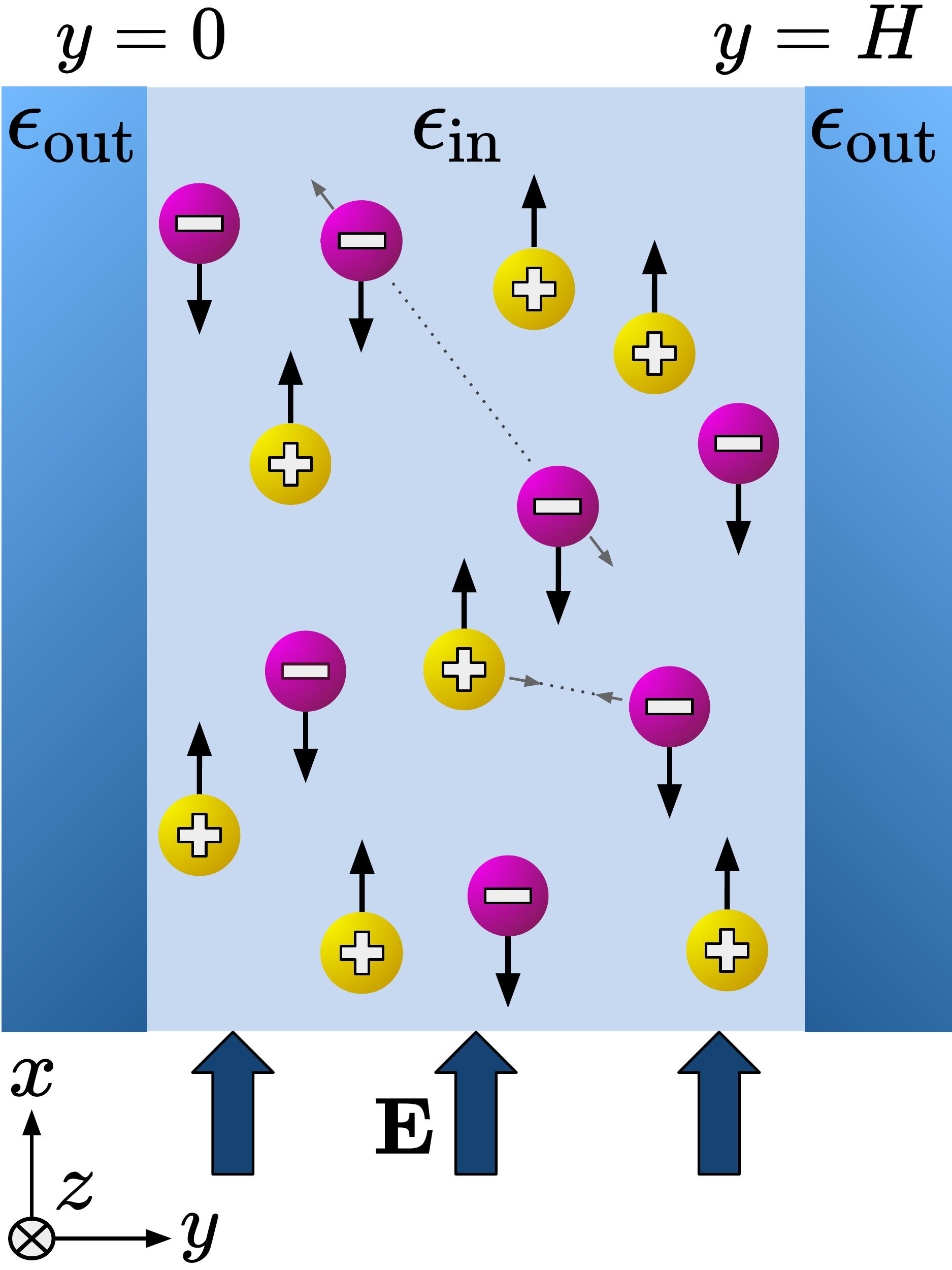}
		\end{minipage}
	\caption{Schematics of an ion channel (left) and the \sm{generic setup of} a  driven electrolyte in flat geometry 
	\sm{studied here} (right). The external electric field $\bm{E} = E \hat{\bm{e}}_x$ drives the positive and negative charges in opposite directions (black arrows) and, in addition, the charges exert electrostatic forces on each other (gray arrows). 
	The system is open along the $x$ and $z$ axes, and it is confined by \sm{\textit{neutral}} plates (with similar boundary conditions) in the $y$ direction.
	} \label{fig:schematic}
\end{figure}

Confined electrolytes 
are highly structured fluids \cite{Perkin2012} that are ubiquitous in various areas of nanotechnology \cite{Siwy2002,schoch2008transport} and  biology \cite{aidley1996ion,Martinac2004}, for instance in biological or synthetic nanopores (see Fig.~\ref{fig:schematic}).
{Recently, there have been experimental observations of force generation in charged solutions which deviate considerably from mean-field predictions~\cite{perez2019surface,stoneAC}.}
It is therefore desirable to understand the generic mechanisms by which correlations in these systems may give rise to fluctuation--induced interactions. 
%
%
%
%
%
%

\begin{figure*}[t]
(a)	\begin{minipage}[c]{.25\linewidth}
		\centering
		\includegraphics[width=\linewidth]{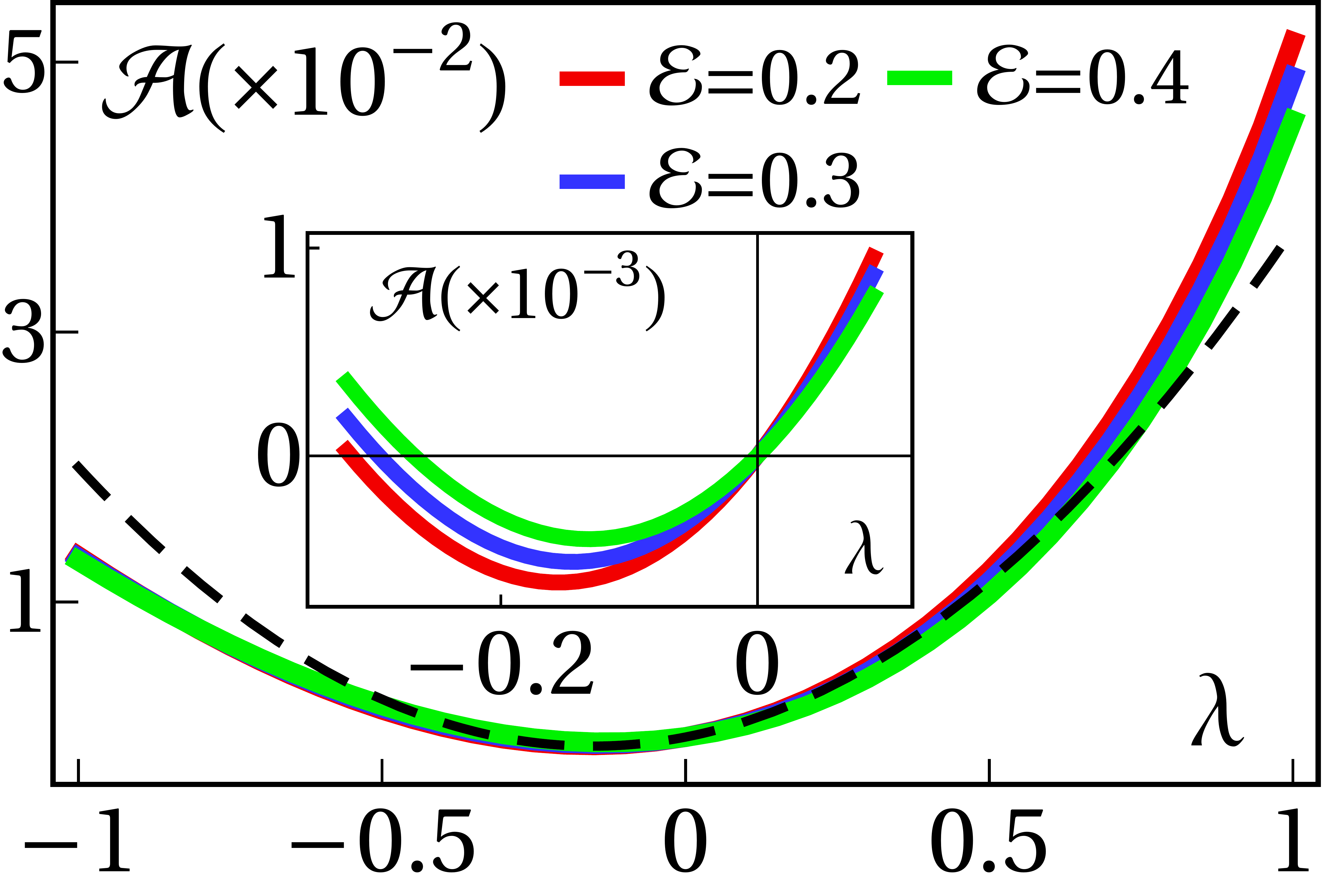}
	\end{minipage} 
	\hskip.5cm
(b)	\begin{minipage}[c]{.25\linewidth}
		\centering
		\includegraphics[width=\linewidth]{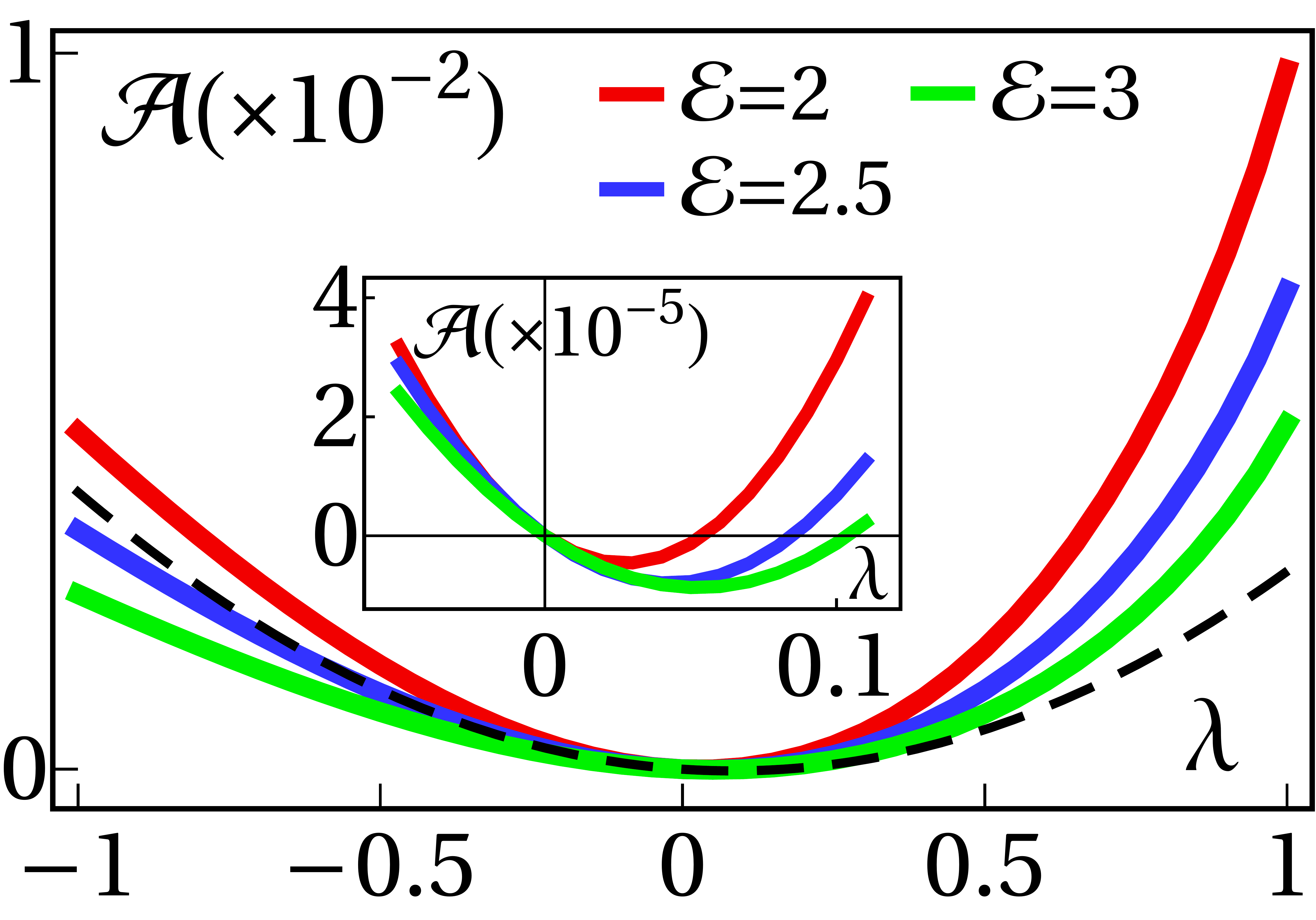}
		\end{minipage} 
			\hskip.5cm
(c)	\begin{minipage}[c]{.25\linewidth}
		\centering
		\includegraphics[width=\linewidth]{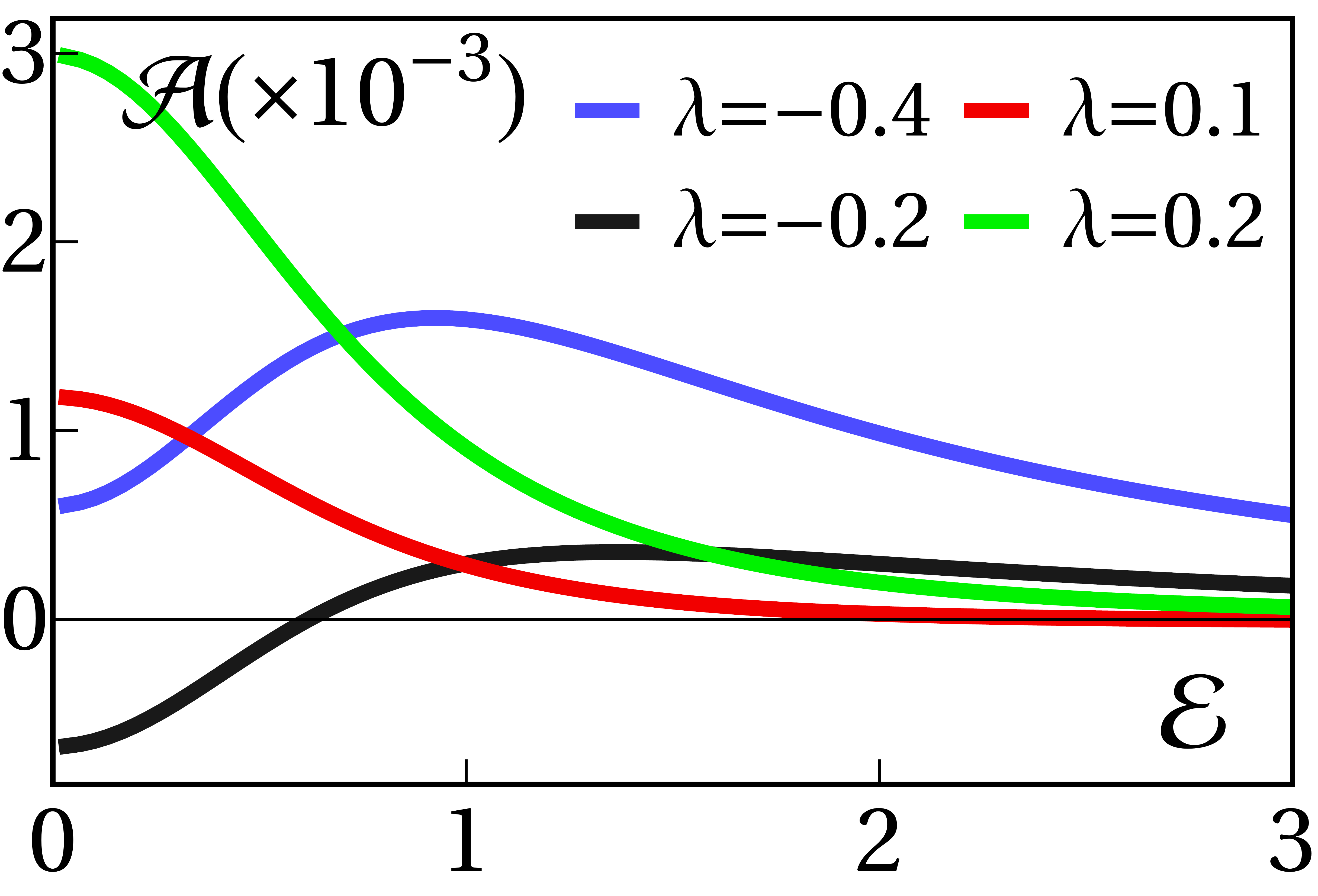}
		\end{minipage}  
	\caption{The FIF amplitude $\calA$ 
	as a function of $\lambda$ for (a) small 
	 and (b) large electric fields. Panel (c) shows $\calA$  as a function of $\calE$ for various values of $\lambda$. 
	The solid lines correspond to the numerical evaluation of Eq.~\eqref{eq:calA}, while the dashed lines in (a) and (b) are, respectively, the approximate forms for $\calE\ll 1$ (independent of $\calE$) and $\calE \gg 1$ (computed for $\calE=3$).
	The asymptotic expressions in Table.~\ref{tab:approx} reveal three distinct behaviors of $\calA$ as a function of $\calE$ (panel c): 
	for $ -0.31 \lesssim \lambda < 0 $, the amplitude $\calA$ is initially negative, crosses to positive values and falls off to zero for large $\calE$; 
	for $0 < \lambda \lesssim 0.17$, the amplitude $\calA$ is initially positive, crosses to negative values before approaching zero from below for large $\calE$; 
	and, finally, for $\lambda \lesssim -0.31$ or $\lambda\gtrsim 0.17$, $\calA$ remains positive when $\calE$ is varied. 
	} \label{fig:Aplots}
\end{figure*}

It is well known that in thermal equilibrium, 
{the correlations in an electrolyte are exponentially screened beyond the Debye length~\cite{israelachvili}. }
Here, we examine the {long-distance behavior} of a strong electrolyte {when it is} driven out of equilibrium by an external electric field, and we show that the anisotropy introduced by the electric field gives rise to power-law correlations and \textit{generically scale invariant} dynamics \cite{grinstein91generic,tauber}. 
Such scale-free correlations have  considerable implications on the dynamics of external boundaries that enclose the electrolyte. 
Using the Maxwell stress tensor, we calculate the FIF that results from confining the electrolyte driven by the external field $\bm{E}=E\hat{\bm{e}}_x$ {parallel to neutral flat boundaries}  
\footnote{{In setups where the electric field is applied perpendicular to the (charged) boundaries, the renormalization of the applied field due to screening effects should be taken into account.}}
(see Fig.~\ref{fig:schematic}); we find that in $d$ spatial dimensions, the {normal} force per unit area of the plates is given by 
\begin{align}   \label{eq:main}
    \frac{F}{S} = 
    -\frac{k_{\rm B} T}{H^d} \, \calE^4 \, \calA(\calE,\lambda) = 
    - \frac{\big(\frac{\epsilon_{\rm in}}{2 S_d} E^2\big)^2}{k_{\rm B} T C_0} \cdot \frac{\calA( \calE ,\lambda)}{C_0 H^d}, 
\end{align}
where $H$ is the separation between the plates. This force depends on the dimensionless parameter $\calE$, which represents the ratio between the electric field (Maxwell) stress and the osmotic pressure of the electrolyte, and on the dielectric contrast $\lambda$ that are defined as follows
\begin{align}   \label{eq:calElambdadef}
    \calE =  
      \left[\frac{\frac{\epsilon_{\rm in}}{2 S_d} E^2}{k_{\rm B} T C_0}\right]^{1/2} , 
     \qquad \text{and} \qquad
     \lambda = \frac{\epsilon_{\rm in}-\epsilon_{\rm out}}{\epsilon_{\rm in} + \epsilon_{\rm out}}.
\end{align}  
In these relations, $\epsilon_{\rm in}$ and $\epsilon_{\rm out}$ are the \sm{permittivities} of the electrolyte and the boundary material, respectively, $C_0$ is the mean concentration (of each charge) in the electrolyte, and $S_d = \frac{2 \pi^{d/2}}{\Gamma(d/2)}$ is the area of the  $d$-dimensional unit sphere. 
The dimensionless amplitude $\calA$ is independent of the applied field for $\calE \ll 1$, implying that the FIF scales as $\sim E^4$ for relatively weak electric fields; for $\calE \gg 1$, on the other hand, one has $\calA \sim 1/\calE^2$, and therefore for large applied electric fields the force scales as $\sim  E^2$ (see Table.~\ref{tab:approx} and Fig.~\ref{fig:Aplots}). 
Intriguingly, the amplitude $\calA$ varies non-monotonically with $\calE$ {(in addition to  $\lambda$)}, and it can also change sign (see Fig.~\ref{fig:Aplots}). 
The \sm{sign change} indicates that the resulting FIF can be tuned to be both repulsive and attractive in the same setup  
\sm{ with symmetric boundary conditions. 
It is worth noting that the force studied here is purely due to fluctuation effects and a change in the sign of this generic long-range FIF is a unique feature that distinguishes our results from, e.g., modifications of critical Casimir forces on introducing additional surface or bulk features that contribute to the force~\cite{dietrich-field1,dietrich-chem1,vasilyev2011critical,dietrich-electro1,pousaneh2014ions}.} 

%
%
%

\emph{Stochastic Density Equations---}
%
We consider a strong simple electrolyte which consists of 
an equal number of cations and anions with charges $\pm Q$ and with equal mobilities $\mu_+=\mu_-=\mu$. 
The individual cations and anions, labeled by $a$, 
move under the combined influence of the external electric field ($\bm{E}$), the electrostatic field of other ions ($-\nabla \phi$), and Brownian motion~\cite{onsagerfuoss} due to thermal noise ($\bm{\eta}_a^\pm$). 
In the overdamped regime, the trajectory $\bm{r}^\pm_a(t)$ of a cation or anion is governed by the Langevin dynamics 
$
    \dot{\bm{r}}_a^\pm(t) = 
    \mu \left(\pm Q\right) \left[ -\nabla\phi\big(\bm{r}_a^\pm(t)\big) + \bm{E} \right] + 
    \sqrt { 2 D } \, \bm{\eta}_a^\pm (t) 
$~\cite{zorkot2016power,demery2016conductivity}. 
The thermal noises $\bm{\eta}_a^\pm$ are independent Gaussian white noises characterized by
$\langle \eta_{a i}^+ (t) \eta_{b j}^+ (t') \rangle = 
\langle \eta_{a i}^- (t) \eta_{b j}^- (t') \rangle =
\delta_{ab} \, \delta_{ij} \, \delta ( t - t' )$
and zero mean ($a$ and $b$ are particle indices and $i$ and $j$ represent vector components).
At this microscopic level, the fluctuation--dissipation relation connects the noise strength $D$ to the mobility $\mu$ through the Einstein relation $\mu = \beta D$, where $\beta = 1/(k_{\rm B} T)$ is the inverse temperature. 
Note that we neglect the hydrodynamic effects of the solvent throughout this work
\footnote{
{The hydrodynamic interactions can become relevant in a concentrated electrolyte (where the hydrodynamic radii of the particles become comparable with interparticle distances) or in extreme confinements. These are not the focus of this letter.} }.

Using the instantaneous number density of each type of charge, which is   defined as  
$ C^\pm(\bm{r},t) = \sum_{a} 
 \delta^d \left( \bm{r} - \bm{r}_a^\pm(t) \right),$
one can express the electrostatic Poisson equation in Gaussian units as
$ -\nabla^2 \phi = \frac{S_d  Q}{\epsilon_{\rm in}}( C^+ - C^-). $
The Dean--Kawasaki approach \cite{dean96langevin,kawasaki1994,ddftreview} then gives the exact dynamics of $C^\pm$ as continuity equations, namely $ \partial_t C^\pm + \nabla\cdot \bm{J}^\pm = 0$, where the stochastic currents of the charges are given by
$\bm{J}^{\pm} = - D \nabla C^\pm \pm \mu C^\pm Q \left(  - \nabla\phi + \bm{E} \right) - \sqrt{2 D C^\pm} \, \bm{\eta}^\pm.$
Here, $\bm{\eta}^\pm$ are uncorrelated Gaussian noise fields with zero averages and $\langle \eta_i^\pm(\bm{r},t) \eta_j^\pm(\bm{r}',t') \rangle =\delta_{ij} \delta^d(\bm{r}-\bm{r}') \delta(t-t')$.

%
The Dean--Kawasaki equations for $C^\pm$ are analytically intractable due to the nonlinear terms 
and the multiplicative noise. 
To avoid these difficulties, we consider the dynamics of the density and charge fluctuations around a state with uniform distribution of the particles, which allows us to linearize the dynamics. 
This simplification remains valid for a dense population of soft particles~\cite{demery2014generalized} and has been used, e.g., to study the conductivity of strong electrolytes \cite{demery2016conductivity}, fluctuations of ionic currents across nanopores \cite{zorkot2016power,zorkot2018nanopore}, and the universal correlations in driven binary mixtures~\cite{poncet2017universal}.
We have also examined the scaling behavior of the nonlinear terms which reveals that these nonlinearities  are irrelevant at the macroscopic level 
(see Ref.~\setcounter{footnote}{100}\footnote{See Supplemental Material at [URL will be inserted by publisher] which contains more details on scaling analysis, correlation functions, and calculation and analysis of the stress tensor in $2$ and $3$ dimensions.} for details). 
%
%
We therefore write the density of each type of charge as $C^\pm = C_0 + \delta C^\pm$ and assume the density fluctuations $\delta C^\pm$ are small compared to the background $C_0$, i.e., 
$\delta C^\pm \ll C_0$.
Introducing the number 
fluctuations $c (\bm{r},t) = \delta C^+  +  \delta C^-$ and the charge 
fluctuations $\rho (\bm{r},t) = \delta C^+  -  \delta C^-$ (in units of $Q$), 
the linearized stochastic equations of $\delta C^\pm$ can be recast as
\begin{align}   
    \partial_t c &= 
    D\nabla^2 c - \mu  Q  \bm{E}\cdot\nabla \rho + \sqrt{4DC_0} \, \eta_c,  \label{eq:linc} \\
    \partial_t \rho &=
    D\nabla^2 \rho - \mu  Q\bm{E}\cdot \nabla c - D \kappa^2 \rho + \sqrt{4DC_0} \, \eta_\rho     \label{eq:linrho}.
\end{align}
Here the Debye screening length $ \kappa^{-1} $ is defined through
$\kappa^2 = 2 S_d C_0 \ellb$
with the ``Bjerrum length’’ 
$\ellb = \beta  Q^2 / \epsilon_{\rm in}$ 
(in $d$ spatial dimensions $\ellb \sim (\text{length})^{d-2}$). 
In addition, the linearized noise correlations read  
$\langle \eta_\rho (\bm{r},t) \eta_\rho (\bm{r}',t') \rangle = \langle \eta_c (\bm{r},t) \eta_c (\bm{r}',t') \rangle = 
-\nabla^2 \delta^d (\bm{r}-\bm{r}')  \delta (t-t') $,
and 
$\eta_{\rho}$ and $\eta_c$  
have zero means and are uncorrelated.
%

%
Without an external electric field ($\bm{E} = 0$), Eqs.~\eqref{eq:linc} and \eqref{eq:linrho} describe the normal diffusion of the number density $c$ and the relaxation of the charge density $\rho$ with the Debye relaxation time $(D\kappa^2)^{-1}$. 
In the presence of the electric field, however, these dynamics become nontrivial as the field couples $\rho$ and $c$.  
In particular, this coupling gives rise to a  charge fluctuation that persists beyond the Debye relaxation time. 
To show this, we focus on the macroscopic limit of Eq.~\eqref{eq:linrho} 
beyond the Debye length- and time-scale $\kappa^{-1}$ and $(D\kappa^2)^{-1}$ which is given by 
\begin{align}   \label{eq:rho(C)}
    \rho \approx -\kappa^{-2}\beta Q\bm{E}\cdot\nabla c, 
\end{align}
where we have made use of the Einstein relation. 
{Eq.~\eqref{eq:rho(C)} gives the charge fluctuations caused by density gradients along the electric field, and it is valid both in  bulk as well as in the presence of boundaries that are parallel to the electric field. } 
The Debye length for typical electrolyte solutions is of order $\kappa^{-1} \sim 1 - 10 \,\rm{nm}$ \cite{israelachvili};  therefore, Eq.~\eqref{eq:rho(C)} applies to {bulk solutions at scales larger than the Debye length, and to confined systems with  boundary separations beyond the screening scale}, e.g., in the case of wet ion channels such as 
mechanosensitive channels \cite{Martinac2004} and synthetic nanopores \cite{Siwy2002}. 
{The charge fluctuations given by Eq.~\eqref{eq:rho(C)} have nontrivial  correlations which, through the Maxwell stress, give rise to long-ranged FIFs as we show in the following. }

Substituting Eq.~\eqref{eq:rho(C)} back into Eq.~\eqref{eq:linc}, we obtain an \textit{anisotropic} diffusion equation for $c$ that reads
\footnote{{We have discarded the noise $\eta_\rho$ when substituting Eq.~\eqref{eq:rho(C)} into Eq.~\eqref{eq:linc}, as it is negligible with respect to $\eta_c$ due to the presence of an additional gradient operator.}}
\begin{equation}       \label{eq:effectivelangevin}
    \partial_t c = 
    D \left(\calE^2 \, \partial_x^2 + \nabla^2 \right) c
    + \sqrt{4 D C_0} \, \eta_c, 
\end{equation}
%
where 
$\calE$ is defined in Eq.~\eqref{eq:calElambdadef}. 
Alternatively, $\calE$ can be expressed  
as $\calE = \mu  Q E/(D\kappa)$, and in this form it encodes the relative deformation of counterion atmospheres in the external field~\cite{onsagerfuoss}.
Note that even though the Einstein relation holds for the microscopic dynamics, 
the anisotropic mismatch between noise and dissipative forces in Eq.~\eqref{eq:effectivelangevin} 
renders the system generically scale invariant \cite{grinstein90conservation, garrido90conservative}, i.e., it automatically gives rise to long-ranged correlations  without tuning parameters. Moreover, this scale invariance also holds for a system where the charge species have different mobilities ($\mu_+ \neq \mu_-$); in that case,  $D$ in Eq.~\eqref{eq:effectivelangevin} is replaced by the (arithmetic) average of the noise strengths of cations and anions 
\footnote{{The difference in mobilities, however, has important implications on the steady electric field in the presence of  oscillatory external driving, see Ref.~\cite{amrei2018oscillating}.}}.
%

\emph{Correlation Functions---}
%
The steady-state correlation functions in the absence of any boundaries can be obtained readily in the Fourier space. 
Defining  
$\langle c (\bm{k},t) c(\bm{k}',t) \rangle \equiv \langle c (\bm{k}) c(\bm{k}') \rangle \equiv  (2\pi)^d \delta^d(\bm{k}+\bm{k}') [ 2 C_0 + c_{\rm bulk}^{(2)}(\bm{k})]$, from Eq.~\eqref{eq:effectivelangevin} one gets 
\begin{align}   \label{eq:c2bulk}
   c^{(2)}_{\rm bulk}(\bm{k})
    = - \frac{ 2 C_0 \calE^2 k_x^2 }{ \calE^2 k_x^2 + k_y^2 + \bm{k}_s^2 },
\end{align}
where we have defined $\bm{k}_s = (k_{s_1}\!=\!k_x, k_{s_2}, \ldots, k_{s_{d-1}}) \in \mathbb{R}^{d-1}$ and $\bm{k}=\bm{k}_s + k_y \hat{\bm{e}}_y$ (i.e., $\bm{k}_s$ is the momentum component parallel to boundaries). 
Equation~\eqref{eq:c2bulk} is the long-distance nonequilibrium correlation due to the external electric field which vanishes for $\calE = 0$, 
and whose $\bm{k}\to 0$ limit is rendered singular by the anisotropy~\cite{tauber}.  
Transforming Eq.~\eqref{eq:c2bulk} back to real space gives 
$c^{(2)}_{\rm bulk} (\bm{r})=-\tfrac{2C_0 \calE^2 (1 - d \tilde{x}^2 / \tilde{r}^2)}{ S_d(\calE^2+1)^{3/2} \tilde{r}^{d}}$
%
where $\tilde{x} = {x}/{\sqrt{\calE^2 +1}}$ and $\tilde{r}$ is obtained from $r$ by substituting $x \to \tilde{x}$.
This expression clearly displays the anisotropy in the correlation function and shows that in $d$ dimensions, the density correlations decrease as $\sim r^{-d}$ with distance. The charge correlations in the bulk can be obtained from 
the density correlations via Eq.~\eqref{eq:rho(C)}.  It is therefore seen that charge fluctuations are long-range correlated as well. 
(Note that the equilibrium part of the correlation functions vanishes  asymptotically at the large distances considered here~\cite{Note101}.) 

We now calculate the steady-state correlations in the presence of boundaries. In particular, we consider a plane parallel geometry with impenetrable 
uncharged boundaries located at $y=0,H$ (Fig.~\ref{fig:schematic}). 
With the Neumann boundary conditions, the solutions of Eq.~\eqref{eq:effectivelangevin} can be constructed by decomposing $c$ and $\eta_c$ onto the cosine modes $\cos( p_{n} y)$ where 
$p_n = \frac{n\pi}{H}$ 
\cite{rohwer2017transfif,barton1989}. This leads to the discretized form of Eq.~\eqref{eq:c2bulk}, viz  
\begin{equation}    \label{eq:c2}
    c^{(2)} ( y , y' ; \bm{k}_s )  =   \frac{-4C_0 k_x^2 \calE^2}{H} 
     \sideset{}{'}\sum_{n=0}^\infty 
     \frac{  \cos ( p_n y) \cos (p_n y')}{ \calE^2 k_x^2 + p_n^2 + \bm{k}_s^2},  
\end{equation}
where we now have 
$\langle c (y; \bm{k}_s) c (y'; \bm{k}'_s) \rangle = (2\pi)^{d-1} \delta^{d-1} (\bm{k}_s + \bm{k}'_s) [ 2C_0 \delta(y-y') + c^{(2)}(y,y';\bm{k}_s)]$ 
and $\sum'$ indicates a $1/2$ factor for the $n=0$ term. 
To obtain the charge correlation function, one needs to make use of Eq.~\eqref{eq:rho(C)} which through Eq.~\eqref{eq:c2} yields
%
\begin{equation}  \label{eq:rho2}
    \rho^{(2)}(y,y'; \bm{k}_s) 
     = 
     \frac{-4 C_0 k_x^4 \calE^4}{ \kappa^2 H }
      \sideset{}{'}\sum_{n=0}^\infty
     \frac{  \cos ( p_n y)  \cos (p_n y') }{   \calE^2 k_x^2 + p_n^2  + \bm{k}_s^2},
\end{equation}
where we have defined
$\langle \rho (y; \bm{k}_s) \rho (y'; \bm{k}'_s) \rangle = (2\pi)^{d-1} \delta^{d-1} (\bm{k}_s + \bm{k}'_s) [ 2C_0 (1 + \calE^2 k_x^2/\kappa^2) \delta(y-y') +  \rho^{(2)}(y,y';\bm{k}_s)]$. 
%
Note that the local contribution $\propto k_x^2 \delta(y-y')$ in charge correlations is due to the deformation of counterion atmospheres 
and it does not contribute to the long-range  forces on the boundaries.  


\emph{Stress Tensor---}
%
%
The nonequilibrium pressure exerted on the boundaries 
can be calculated using the noise-averaged Maxwell stress 
$\langle \sigma_{ij} \rangle= 
\frac{\epsilon_{\rm in}}{2 S_d} \left( 2 \langle \nabla_i \phi \, \nabla_j \phi \rangle
- \delta_{ij} \langle (\nabla\phi)^2 \rangle \right)$
\footnote{Using the Maxwell stress in this nonequilibrium setting is justified as it derives from the electrostatic force density $\nabla \cdot \bm{\sigma}=-\rho\nabla\phi$ \cite{woodson1968electromechanical} in a similar way as the nonequilibrium Irving--Kirkwood formula is constructed from microscopic forces \cite{irving1950hydro,kruger2018neqstress}}. 
We assume no free charges on the boundaries, and hence the electrostatic boundary conditions are given by the continuity of the tangential electric field and the normal  displacement field at both plates.
Using image charges~\cite{jackson}, 
the electric potential satisfying these conditions is obtained  as
$\phi (\bm{r}) = \sum\limits_{n \in \mathbb{Z}}
\int_{\bm{r}'} \, \tfrac{\lambda^n  \rho(\bm{r}')  Q }{ \epsilon_{\rm in} (d-2)|\bm{r}-\bm{r}'_n|^{d-2}}$
\footnote{For $d=2$, the potential is given by the logarithmic Coulomb form, while the Maxwell stress formula remains unchanged.}
where $n$ indexes the image located at 
$\bm{r}'_n$ (obtained from $\bm{r}'$ by substituting $y' \to y'_n = 2nH \pm y'$), 
and $\lambda$ [as defined in Eq.~\eqref{eq:calElambdadef}]
represents the ratio of the successive image charges. 

On substituting $\phi(\bm{r})$ into the Maxwell stress, performing the summation over image charges, and using Eq.~\eqref{eq:rho2}, we arrive at Eq.~\eqref{eq:main} for the normal force on the plate at $y=H$
\footnote{The force exerted on the $y=0$ boundary has the same magnitude and is in the opposite direction.}. 
The stress amplitude in Eq.~\eqref{eq:main} reads   
$\calA=\lambda \int_{\bm{\nu}_s} \lbrace \sum\limits_{n=0}^\infty\!\!{\vphantom{\sum}}' \calR_n (\lambda,\nu_s) g(\calE,n,\bm{\nu}_s) -  \int_{0}^\infty \dif n \, g(\calE,n,\bm{\nu}_s) \rbrace$, 
where we have defined $\calR_{2k}= \calY_{-}$, $\calR_{2k+1}=\calY_{+}$, $\calY_{\pm}\equiv  \left(\frac{e^{\pi\nu_s}\pm 1}{e^{\pi\nu_s}\pm\lambda}\right)^2$, 
and 
$g = \tfrac{  2^{1-d} \nu_x^4 \bm{\nu}_s^2}{(\calE^2\nu_x^2 + n^2+ \bm{\nu}_s^2)(n^2 + \bm{\nu}_s^2 )^2}$.
For the $d=3$ case, we find~\cite{Note101} 
\begin{widetext}
\begin{align}       \label{eq:calA}
     &\calA(\calE,\lambda) = \frac{\lambda \zeta(3)}{16 \pi} \frac{\calE^2 + 2}{ \calE^4 (\calE^2+1)^{3/2}} 
    +\frac{\calE^2 - 4}{32\pi \calE^4} \left[ (\lambda - \frac{1}{\lambda}) \Li_2(\lambda^2) +\frac{1}{2} (\lambda+\frac{1}{\lambda}) \Li_3(\lambda^2) \right] + \frac{3 \Li_3(\lambda^2)}{32\pi\calE^2} \nonumber 
    \\
    &\qquad + \frac{\lambda \pi}{16} \int_0^{2\pi} \dif\theta \int_0^\infty \nu_s^2 \dif \nu_s \frac{ \left[\calY_-(\lambda,\nu_s) -1 \right] \coth( \frac{\pi\nu_s}{2} \sqrt{\calE^2\cos^2\theta +1})  +  \left[\calY_+ (\lambda,\nu_s) -1\right] \tanh(\frac{\pi\nu_s}{2}\sqrt{\calE^2 \cos^2\theta +1})}{ \calE^4 \sqrt{\calE^2 \cos^2 \theta+1}} 
\end{align}
\end{widetext}
where ${\Li}_n(z) = \sum\limits_{k=1}^\infty \dfrac{z^k}{k^n}$ is the polylogarithm function. 
Figure~\ref{fig:Aplots} shows the variation of $\calA$ as a function of $\lambda$ [Fig. \ref{fig:Aplots}(a) and Fig. \ref{fig:Aplots}(b)] and $\calE$ [Fig. \ref{fig:Aplots}(c)] in $d=3$ dimensions, as obtained numerically from Eq.~\eqref{eq:calA}. 
In Table~\ref{tab:approx}, we also summarize the approximate forms for $\calA$ in $d=3$ obtained from the suitable asymptotic expansion of Eq.~\eqref{eq:calA} in each regime~\cite{Note101}. These approximate forms are shown in Fig.~\ref{fig:Aplots}(a) and Fig.~\ref{fig:Aplots}(b) as dashed lines. 
%

%
\begin{table}[b]
    \centering
    \begin{ruledtabular}
         \begin{tabular}{c  c  c}
          &   $\calE \ll 1$ 
          & $\calE \gg 1$     \\ \hline
         {$\lambda \!\ll \! 1$} \rule{0pt}{5ex}   
         & ${ \dfrac{9\left( 4 \zeta(5) + 1 \right)}{512 \pi} \lambda^2 \!+\!  \dfrac{9\left( 4 \zeta(5)-1 \right)}{1024 \pi } \lambda}$ 
         & $\dfrac{ \lambda (6\lambda-1) }{64 \pi}\calE^{-2}  $\\[.3cm]
         {$\lambda \!=\! 1$} 
         & $\dfrac{9 \, \zeta (3) }{64 \pi}$ 
         & $\dfrac{\zeta (3)  }{ 8 \pi } \calE^{-2} $ \\[.3cm]
          $\lambda \!=\! -1$ 
          & $\dfrac{9 \zeta (3) }{256 \pi}$ 
          & $\dfrac{\zeta (3)  }{16 \pi }\calE^{-2}$ \\
        \end{tabular}
        \end{ruledtabular}
        \caption{Leading order approximations of the dimensionless stress amplitude $\calA (\calE,\lambda)$ in $d=3$, obtained from expanding Eq.~\eqref{eq:calA} 
        (corrections are $\mathcal{O}(\calE^2)$ for $\calE \ll 1$,  and $\mathcal{O}(\calE^{-4})$ for $\calE \gg 1$). 
        The relevant values of the Riemann zeta function are $\zeta(5) \approx 1.04$ and $\zeta(3) \approx 1.20 $.  
        (For derivations as well as asymptotic expressions in $d=2$, see Ref.~\cite{Note101}.)
        \label{tab:approx}}
    \end{table} 

From Eq.~\eqref{eq:calA} and the asymptotic forms in  Table.~\ref{tab:approx}, we note that the FIF exhibits two different regimes for weak and strong applied electric fields: for weak fields ($\calE \ll 1$), the force scales as $E^4$, and it is proportional to the inverse temperature ($\propto \beta$) and the inverse average density squared ($\propto 1/C_0^2$); 
for strong fields ($\calE \gg 1$), the force scales as $E^2$, and it is  proportional to the inverse average density ($\propto 1/C_0$) and becomes independent of temperature. (Note that the second line in Eq.~\eqref{eq:calA} is a subleading correction for $\calE \gg 1$.)
Fig.~\ref{fig:Aplots} also shows that the sign of the force amplitude can change with the applied electric field: 
for $\lambda \ll 1$ (i.e. small dielectric contrast), 
the amplitude $\calA$ can become negative which, remarkably, indicates a repulsive 
FIF between boundaries that enclose the driven electrolyte.
%
For moderate values of the dielectric contrast $\lambda$, on the other hand, the force remains attractive with a positive amplitude.

\textit{Concluding Remarks---} 
We showed that the effective anisotropy introduced by the external electric field renders the steady-state density and charge fluctuations in a driven electrolyte long-range correlated, in contrast to the screened correlations in thermal equilibrium. 
These nonequilibrium correlations give rise to long-range FIFs on external objects and boundaries immersed in the driven electrolyte.  
For neutral boundaries parallel with the external field \sm{(Fig.~\ref{fig:schematic})}, our results show that the  force  
varies non-monotonically with the electric field and the dielectric contrast;  notably, \sm{with symmetric boundaries,} the \sm{long-range} FIF 
can be tuned to be attractive or repulsive with different amplitudes by varying the relevant parameters. 
We note the present setup gives an independent mechanism for tuning nonequilibrium FIFs from the one in Ref.~\cite{deannoneqtune2016}, and investigating the interplay of the two effects forms an interesting direction for future studies.

This work highlights a generic mechanism through which nonequilibrium fluctuations can give rise to long-range forces in driven charged systems. 
\sm{ A similar mechanism may also be relevant for concentrated electrolytes where it has been suggested 
that the ``defects" are the main  charge carriers~\cite{perez2017scaling}, and it may also be applicable to ionic liquids where mainly the ``free ions" participate in conduction and screening processes~\cite{lee2015room,kornyshev}.}
Long-ranged forces have recently been observed in a number of experimental settings~\cite{perez2019surface,stoneAC} where oscillatory electric fields are applied to charged solutions.  
\sm{Our preliminary analysis shows that the FIF introduced here can help to understand the experimentally observed dynamical features;  
we plan to extend our results to include time-dependent electric fields and different boundary conditions.}  
%
%
In addition, the model used here relies on the linearization of the stochastic dynamics which is applicable to strong electrolytes far from phase transitions and where the Gaussian scalings hold~\cite{Note101}.
We plan to perform a more rigorous treatment of the nonlinearities using renormalization group (RG) techniques which have recently been applied to similar dynamics in the context of chemotaxis \cite{chemoPRR}.

\textit{Acknowledgments --} The authors thank A.~Gambassi, A.~Macio{\l}ek, S.~Dietrich, M.~Bier, and M.~Gross for pointing out the difference in  screening of the external field with perpendicular setups  
and for bringing the references on critical Casimir force to our attention. 
\sm{We also thank Susan Perkin for helpful discussions.}
S.M. acknowledges the support of the Clarendon Fund and St John's College Kendrew Scholarship from the University of Oxford.
This work was supported by the Max Planck Society.

\bibliography{new_bib}

\end{document}